\RequirePackage{lineno}
\documentclass[prl,twocolumn,preprintnumbers,tightenlines,superscriptaddress]{revtex4}
\usepackage{graphicx}
\usepackage{amsmath} 
\usepackage{xcolor}
\usepackage{tikz}
\usetikzlibrary{positioning}
\usetikzlibrary{decorations.pathmorphing}
\usetikzlibrary{decorations.markings}
\usetikzlibrary{patterns}

\usepackage[hidelinks]{hyperref}
\usepackage{xspace}

\usepackage[]{changes}

\definecolor{darkblue}{rgb}{0, 0, 0.72}

\definechangesauthor[name={Yanou},    color=red     ]{yc}
\definechangesauthor[name={Yue},      color=blue    ]{yz}
\definechangesauthor[name={Lina},     color=darkblue]{lmn}
\definechangesauthor[name={Yun-Tse},  color=cyan    ]{ytt}
\definechangesauthor[name={Josh},     color=magenta ]{jb}
\definechangesauthor[name={Gianluca}, color=orange  ]{gp}

\newcommand{\fakesubsection}[1]{\textbf{\textit{#1.}}\hspace{1em}}

\newcommand{\Cherenkov}{Cherenkov\xspace}
\newcommand{\kt}{kiloton\xspace}
\newcommand{\GEANT}{\textsc{Geant4}\xspace}
\newcommand{\GENIE}{GENIE\xspace}
\newcommand{\LArSoft}{LArSoft\xspace}
\newcommand{\PYTHIA}{\textsc{Pythia}\xspace}
\newcommand{\SolTrack}{\textit{SolTrack}\xspace}

\newcommand{\LetterTitle}%
{Prospects for Detecting Boosted Dark Matter in DUNE through Hadronic Interactions}

\newcommand{\supplMat}{\textit{Supplementary Material}}

\newcommand{\BDMmass}{\ensuremath{m_{\chi}}}

\newcommand{\argonEfficiency}{\ensuremath{\epsilon_{\textnormal{Ar}}}}
\newcommand{\referenceXsecBDMAr}{\ensuremath{\sigma_{\chi\textnormal{Ar}}}}
\newcommand{\referenceBDMflux}{\ensuremath{\Phi}}
\newcommand{\expectedBDMsignal}{\ensuremath{s}}
\newcommand{\expectedBackground}{\ensuremath{b}}

\begin{document}
\tikzset{
	scalar/.style={dashed},
	scalar-ch/.style={dashed,postaction={decorate},decoration={markings,mark=at
			position .55 with {\arrow{>}}}},
	fermion/.style={postaction={decorate}, decoration={markings,mark=at
			position .55 with {\arrow{>}}}},
	gauge/.style={decorate, decoration={snake,segment length=0.2cm}},
	gauge-na/.style={decorate, decoration={coil,amplitude=4pt, segment
			length=5pt}}
}

\preprint{PITT PACC 1910}
\preprint{SLAC-PUB-17486}

\title{\LetterTitle} 

\author{Joshua~Berger}
  \email{Joshua.Berger@colostate.edu}
  \affiliation{Colorado State University, Fort Collins, CO 80526, USA}
  \affiliation{Pittsburgh Particle Physics, Astrophysics, and Cosmology Center, Department of Physics and Astronomy, University of Pittsburgh, Pittsburgh, PA 15260, USA}
\author{Yanou~Cui}%
  \email{yanou.cui@ucr.edu}
  \affiliation{Department of Physics and Astronomy, University of California Riverside, CA 92521, USA}%
\author{Mathew~Graham}
  \email{mgraham@slac.stanford.edu}
  \affiliation{SLAC National Accelerator Laboratory, Menlo Park, CA 94025 USA}
\author{Lina~Necib}%
  \email{lnecib@caltech.edu}
  \affiliation{Walter Burke Institute for Theoretical Physics, California Institute of Technology, Pasadena, CA 91125, USA}%
\author{Gianluca~Petrillo}
  \email{petrillo@slac.stanford.edu}
  \affiliation{SLAC National Accelerator Laboratory, Menlo Park, CA 94025 USA}
\author{Dane~Stocks}
  \email{dstocks@stanford.edu}
  \affiliation{SLAC National Accelerator Laboratory, Menlo Park, CA 94025 USA}
  \affiliation{Stanford University, Stanford, CA 94305, USA}
\author{Yun-Tse~Tsai}%
  \email{yuntse@slac.stanford.edu}
  \affiliation{SLAC National Accelerator Laboratory, Menlo Park, CA 94025 USA}
\author{Yue~Zhao}%
  \email{zhaoyue@physics.utah.edu}
  \affiliation{Department of Physics and Astronomy, University of Utah, Salt Lake City, UT, 84112, USA}%

\date{\today}

\begin{abstract}
	Boosted dark matter (BDM) is a well-motivated class of dark matter (DM) candidates
	in which a small component of DM is relativistic at the present time.
	We lay the foundation for BDM searches via hadronic interactions
	in large liquid-argon time-projection chambers (LArTPCs), such as DUNE.
	We investigate BDM-nucleus scattering in detail by developing new event generation techniques with a parameterized detector simulation.
	We study the discovery potential in a DUNE-like experiment
	using the low threshold and directionality of
	hadron detection in LArTPCs and compare with other experiments. 
\end{abstract}

\maketitle


%
%
%

%
%
%

\fakesubsection{\label{sec:level1}Introduction}
Despite the overwhelming gravitational evidence for the existence of dark
matter (DM), its microscopic nature remains a profound puzzle.
A leading DM paradigm is that of Weakly Interacting Massive Particles (WIMP),
consisting of a single species of deeply non-relativistic particles. 
%
%
Over the past few decades, however,
DM detection experiments have excluded large swaths of the
parameter space for WIMPs~\cite{Agnese:2015nto,Amole:2016pye,Akerib:2016vxi,
Cui:2017nnn,Amole:2017dex,Aprile:2018dbl,Xia:2018qgs,Akerib:2018hck},
motivating serious consideration of non-minimal models and alternative candidates.

%
%
%
In a class of models beyond the minimal WIMP scenario, a small relativistic
component of DM, boosted dark matter
(BDM)~\cite{Huang:2013xfa,Agashe:2014yua, Berger:2014sqa},
is produced at the present time and can be detected via its interactions with the Standard
Model (SM) particles. The detection of BDM could be a smoking gun for
DM discovery in cases where the dominant component of DM is hard to detect, yet
it requires new experimental strategies beyond the current DM searches.

%
%
BDM may originate from scenarios of dark sectors,
with multiple components of DM or with non-minimal stabilization mechanisms,
such as semi-annihilating DM~\cite{DEramo:2010keq,Berger:2014sqa},
self-annihilating DM~\cite{Carlson:1992fn,Hochberg:2014dra},
decaying DM~\cite{Kopp:2015bfa,Cui:2017ytb},
DM induced nucleon decay~\cite{Davoudiasl:2010am,Huang:2013xfa},
or cosmic ray acceleration~\cite{Bringmann:2018cvk,Ema:2018bih,Cappiello:2019qsw,Dent:2019krz}.
A minimal two-component scenario described in the original works~\cite{Huang:2013xfa,Agashe:2014yua, Berger:2014sqa}
includes a cold component $\psi$ as the dominant component of DM
with very small scattering cross sections with the SM particles,
and a relativistic, less massive secondary component $\chi$
produced by the annihilation of $\psi$,
that effectively interacts with the SM particles.
%
%
Thermal freeze-out via processes such as  $\psi\overline{\psi}\rightarrow\chi\overline{\chi}$ annihilation
in the early Universe may determine DM relic abundance
as a new realization of WIMP miracle~\cite{Belanger:2011ww,Agashe:2014yua,Berger:2014sqa}.
Meanwhile, present-day annihilation in DM-concentrated regions,
such as the Galactic Center (GC) or the Sun,
generates BDM $\chi$ that can be detected via their interaction with electrons or hadrons.
%
%
%
%

%
%
The phenomenology of BDM features a relatively small flux and typically
(semi-)relativistic outgoing SM particles upon BDM-SM particle scattering.
As conventional DM direct detection experiments focus on the detection of low-energy
recoils in a detector mass
up to a few tons,
they generally do not have
the best sensitivity for BDM searches (with the exception of very low mass DM,
see \cite{Cherry:2015oca,Cui:2017ytb,Giudice:2017zke,McKeen:2018pbb}). On the
other hand, massive neutrino detectors, sensitive to energetic SM
particles, stand out as ideal facilities.

%
%
Experimentally, BDM can be observed in interactions with electrons
or with nuclei.
In order to cover all possibilities, both kinds of interactions should be studied.
While previous work has extensively studied the parameter space probed by BDM models
when $\chi$ scatters off electrons~\cite{Agashe:2014yua,KONG2015256,Necib:2016aez,
Alhazmi:2016qcs,Kim:2016zjx,Kachulis:2017nci,PhysRevD.98.075027,PhysRevD.100.035033,Kim:2020aa},
BDM detection via interaction with nuclei, which is complicated by nuclear effects, is much less studied. Nuclear scattering is, however, the dominant process in numerous well-motivated models~\cite{
	Nelson:1989fx,He:1989mi,Carone:1994aa,Bailey:1994qv,FileviezPerez:2010gw,
	Graesser:2011vj,Batell:2014yra,Tulin:2014tya,Dobrescu:2014fca},
making this interaction a potential discovery channel. It is possible that leptonic interactions are not even present at all.
%
%
%
%
%
%
%
%

We expect the main background for BDM interactions in a detector to be from
atmospheric neutrinos interacting via the neutral current,
leaving activity in an energy range similar to the signal.
Unlike this background,
all the BDM signal comes from a single source, and
the signal contribution can therefore be enhanced
by selecting events aligning with the source's location.
This is done by selecting events based on the total momentum of all the
detectable particles produced in the interaction.

The exploration and development of the novel technology of liquid-argon time-projection chambers (LArTPCs)
as neutrino detectors has ramped up in the last decade,
and will culminate in the upcoming next generation neutrino facility,
Deep Underground Neutrino Experiment
(DUNE)~\cite{DUNETDRvol1,DUNETDRvol2,DUNETDRvol3,DUNETDRvol4,Cherry:2015oca,Necib:2016aez,Alhazmi:2016qcs,Kim:2016zjx,Ema:2018bih,Chatterjee:2018mej,Grossman:2019aa,PhysRevD.100.035033,Kim:2020aa,Abi:2020kei,DeRoeck:2020ntj}.
Recent studies have shown that search for BDM via
interactions with electrons would benefit from DUNE's excellent
particle identification~\cite{Arneodo:2006ug,Anderson:2011ce,Antonello:2013gut,Acciarri:2014gev,Necib:2016aez,
Alhazmi:2016qcs, Kim:2016zjx,Acciarri:2016sli,Palamara:2016uqu,Acciarri_2017,
Chatterjee:2018mej,Acciarri:2018ahy,PhysRevD.99.091102,PhysRevD.99.092001,Castiglioni:2020tsu,
Adams_2020,Abi:2020mwi}.
LArTPCs are, however, expected to most significantly
improve the sensitivity to BDM in the \textit{hadronic channels}.
The accessible kinematic range for hadrons in water \Cherenkov detectors
is limited by the \Cherenkov threshold (a momentum of $1.07~{\rm GeV}$ for
protons) and, in the case of inelastic scattering, by the quality
of the reconstruction of overlapping rings~\cite{Berger:2014sqa,Fechner:2009aa}.
Massive detectors based on liquid scintillators do not provide directionality,
and segmented liquid scintillators, which can offer the
details of an event such as directionality, have either a small
volume or a relatively coarse granularity, due to cost. LArTPCs have millimeter resolution,
leading to a low detection threshold of hadrons and an ability to
reconstruct recoil direction. They are scalable and have excellent capabilities
in calorimetry and thereby particle identification. These features combine to
alleviate the limitations of current experiments.
Multi-\kt LArTPC experiments like DUNE, which features a fiducial mass of 40 kilotons,
therefore hold great potential for BDM searches.

Liquid argon, a dense fluid with moderately large nuclei, is a consummate target candidate
of BDM detection through hadronic scattering, granting higher interaction rates. 
%
%
Conversely, the nuclear effects of argon,
and in particular the propagation and interaction through the nucleus
(known as final-state interactions, or FSI)
of the hadrons produced in the BDM-nucleon interactions,
alter the kinematics of the particles revealed in the detector~\cite{DUNETDRvol2,Andreopoulos:2009rq},
jeopardizing the reconstruction, and in particular that of the direction, of the signal candidates.

%
%

In this \textit{Letter},
we study the observed hadronic signatures of BDM in LArTPC detectors
evaluating the nuclear effects using a novel Monte Carlo (MC)-based analysis,
and obtain the search sensitivity taking into consideration
the atmospheric neutrino background.


%
%

\fakesubsection{\label{sec:bdm}BDM Model}
%
We consider the following representative BDM model as a benchmark for our
study~\cite{Berger:2014sqa}.
The model consists of two components of DM. The dominant component, $\psi$,
scatters off of hydrogen in the Sun,
gets captured and builds up,
then annihilates into the relativistic lighter component, $\chi$, i.e.\ the BDM:
\begin{equation}
\psi + \overline{\psi} \to \chi + \overline{\chi}.
\end{equation}
The modeling of this annihilation is not particularly relevant to the
phenomenology at hand, but we assume that this is the dominant
annihilation process for $\psi$.
As we discuss shortly, so long as the
annihilation cross section is sufficiently large, it will not enter into the
determination of the BDM flux. Although annihilation is also possible in the Galactic Center,
for a broad range of parameters the flux from the Sun will dominate over that from
the Galactic Center.  The large solar flux makes it possible to have observable signals with scattering cross
sections of weak scale size or even smaller.
%
%
%
%
%
%
%
%
%
%
%
%
%
%

Equilibrium between DM capture and DM annihilation is generically reached in the solar
core~\cite{Huang:2013xfa, Berger:2014sqa},
eliminating the parametric dependence on the DM annihilation cross section.
%
As a minimal assumption,
we do not introduce leptonic interactions,
although the model can accommodate both leptonic and hadronic couplings as independent interactions.
More details of this model
as well as of semi-annihilation scenarios can be found in~\cite{Berger:2014sqa}.%

%
%
\fakesubsection{Hadronic interactions}
The BDM $\chi$ produced in the above process then emerges from the Sun at high
velocity and scatters off of nuclei in the detector, via a process of the form
\begin{equation}
\chi + N \to \chi + X,
\end{equation}
where $N$ is a nucleon and $X$ is any number of hadrons. 
Hadronic DM interactions share similarities with neutral current neutrino scattering,
which makes it natural to perform simulations in the framework of
neutrino MC software suites, such as \GENIE~%
\cite{Andreopoulos:2009rq,Andreopoulos:2015wxa,Berger:2018}.
For this analysis we introduce in \GENIE a BDM module~\cite{Berger:2018}
to perform all of the cross section calculations and event generation.
\GENIE simulates several nuclear effects,
such as nucleon motion, Pauli blocking, and final state interactions of hadrons
escaping the nuclear remnant after scattering.  It further includes parton
distributions, fragmentation, and hadronization in deep inelastic scattering,
with some corrections to deal with the relatively low energy regime of
interest.
The $hA$ final state model~\cite{Andreopoulos:2015wxa,Merenyi:1992gf,Ransome:2005vb}
is employed to model nuclear effects,
though it can be changed to compare with other models.
In the energy regime relevant to LArTPC neutrino detectors, $E_{\nu}\gtrsim 100\,\textnormal{MeV}$,
coherent nuclear scattering is highly suppressed and scattering is dominantly
off nucleons, that become unbound from the nucleus, or electrons, that have
negligible binding energies compared to the momentum transfer.

We include elastic scattering, yielding a recoiling nucleon,
and deep inelastic scattering (DIS), yielding multi-hadron final states,
in our modeling,
while conservatively neglecting resonant inelastic
scattering processes during which an excited baryon is produced and
decays~\footnote{These processes can be comparable in size to the elastic and
deep inelastic processes that we consider here, but they are significantly more
challenging to calculate and simulate.}.
The diagrams of all the three processes are shown in Fig.~\ref{fig:processes}
in the {\supplMat}.
Elastic scattering off nucleons can be described by an
axial form factor.  As for neutrino scattering in \GENIE, the
axial form factor is assumed to have a dipole form. The
normalization of this form factor is given by the spin form factors, which are
currently best determined by lattice QCD calculations~\cite{Alexandrou:2017hac}.
The hadronic component of the DIS scattering cross section is described by a hadronic tensor,
which depends on parton distribution functions (PDF).
\GENIE uses a PDF that includes corrections for the relatively low energy regime of interest.
The fragmentation and hadronization of the final state
depends on the invariant mass of the final state hadronic system.
At low invariant masses, an empirical model is used~\cite{Koba:1972ng},
in which we assume that DM scattering is similar to neutrino scattering.
At high invariant masses, a model based on \PYTHIA~\cite{Sjostrand:2000wi} is used.
Further details can be found in the {\supplMat}.

\fakesubsection{Analysis strategy}
For concreteness, we focus on a benchmark in which both components of DM are scalars.
Both are required to interact with quarks
in order to enable solar capture for the heavy component,
and terrestrial detection for the light component.
The interactions with the SM particles are mediated by a spin-1 vector boson, $Z^\prime$,
with a gauge coupling $g_{Z^\prime}$.
We assume that the quark current is axial. Both
DM species, as well as the SM quarks, have a charge under this interaction,
which is a free parameter of the model. Without loss of generality we take the
BDM charge $Q_\chi = 1$. As a simple benchmark, we
take the quark charges $Q_f = 1$ for all quark flavors, and consider
$m_{Z^\prime} = 1~{\rm GeV}$.
For $m_{Z^\prime} \gtrsim 1~{\rm GeV}$, the effect of the $Z^\prime$ on the BDM
scattering kinematics is small. We leave the heavy DM charge $Q_\psi$ and gauge
coupling $g_{Z^\prime}$, as well as
the masses of the DM species $m_\psi$, \BDMmass{} as free parameters.  Note that
the lighter DM emerges from the Sun with a Lorentz boost
\begin{equation}
\gamma = \frac{m_\psi}{\BDMmass}.
\end{equation}
Even with a mild hierarchy of masses, the velocity of $\chi$ from the Sun can be
much larger than that of the virialized DM in the Solar System, such that
$\chi$ can escape and reach Earth as BDM.

Within this model, we determine the flux of BDM $\chi$ through a detector on
Earth.  The flux depends on three sequential processes: capture,
annihilation, and rescattering. 
The DM capture rate in the model considered has been determined
in~\cite{Berger:2014sqa} and we use a similar calculation.
Given a heavy DM mass, $m_\psi \gtrsim 4~{\rm GeV}$, and a large enough
annihilation cross section, $(\sigma v)_{\rm ann} \gtrsim 3 \times 10^{26}~
{\rm cm^3}/ {\rm sec}$, DM loss through annihilation and DM gain through capture
reach the equilibrium within the lifetime of the Sun over the entire parameter
space to which a multi-\kt LArTPC is sensitive. In this case, BDM flux is
simply determined by the DM capture rate.
In addition, direct detection experiments and Super-Kamiokande
exclude the parameter region in which BDM rescatters as it exits the
Sun~\cite{Berger:2014sqa}. Rescattering is thus negligible for the parameter
range of interest for this study, leading to a nearly monoenergetic flux of
$\chi$.
Combining the above processes, we find that the magnitude of the flux
\begin{equation}
\label{eq:flux}
\Phi = \frac{C}{4 \pi D^2},
\end{equation}
where $C$ is the $\psi$ capture rate, proportional to $g_{Z^\prime}^4$,
and $D$ is the distance from the Sun to the
Earth, i.e.\ 1~AU.
%
%

We scan over the parameter space of four BDM masses $m_\chi$ in the range of
(5 -- 40) GeV and three boost factors $\gamma= 1.1, 1.5, 10.0$,
while probing the coupling constant 
$g_{Z^\prime}$. For a mass $m_\psi$ below 5~GeV, evaporation
of captured dark matter would lead to drastically reduced flux on the Earth,
while above 40 GeV, the DM mass no longer has a significant effect on
the detection efficiency. Our three choices of $\gamma$, in order, represent the
benchmark cases
where the BDM-hadron interaction is all elastic scattering,
a mixture of elastic and inelastic scattering, and mostly inelastic scattering.

We use GENIE to simulate the BDM signals and the atmospheric neutrino
background, where the BDM signal simulation, discussed above,
is developed for this analysis.
The direction of the Sun with respect to the detector,
evaluated based on the \SolTrack{} package~\cite{SolTrack}
and on the geographical coordinates of DUNE~\cite{Acciarri:2016crz} as an example,
is encoded in the samples.
Based on the Bartol atmospheric neutrino flux~\cite{PhysRevD.70.023006} at Soudan,
the atmospheric neutrino samples include neutral-current (NC) neutrino events
and events where $\nu_{\tau}$ interacts with the detector target via charged current (CC)
and the outgoing $\tau$ leptons decay into hadrons.
The rest of the CC neutrino interactions is not included in the background sample,
as we assume with the information offered by LArTPCs,
it can be efficiently rejected by discarding the events
which contain muons or electrons as final state particles~\footnote{
Preliminary studies in a large LArTPC experiment already showed
great separation power between muons and charged pions~\cite{Acciarri_2017},
and further improvements are being pursued~\cite{Castiglioni:2020tsu}.
We expect the residual inefficiencies to be small enough not to be significant in our study,
and we consequently assume 100\% rejection of muons in this analysis.
}.

\begin{figure}
\centering
\includegraphics[width=0.45\textwidth]{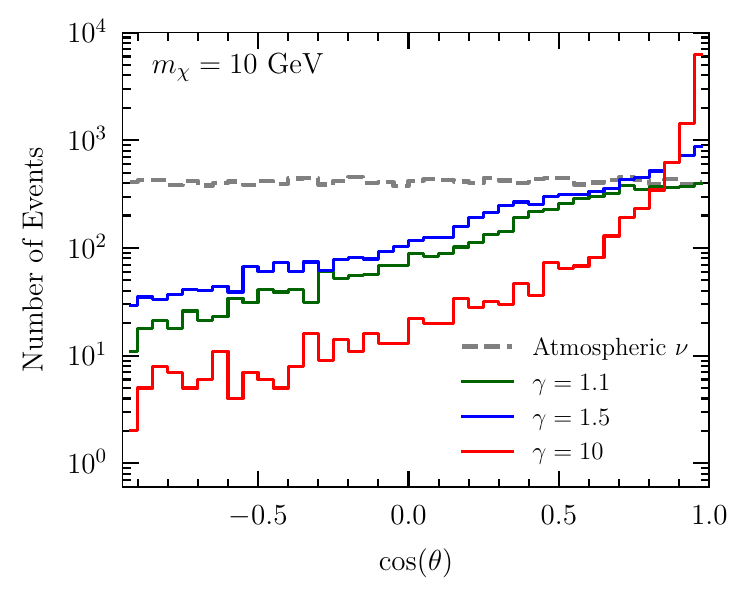}
\caption{The angular distribution $\cos\theta$ of the hadronic BDM signal and the
background with respect to the Sun, which we use to select a sample with optimal
signal-to-background ratio. The three BDM signal samples with the mass $m_\chi$
of 10~GeV but different energies (boost factors $\gamma$) are presented here,
together with the background sample. Nonetheless, the BDM samples with the same
boost factor share the same feature, regardless the BDM mass $m_\chi$. All the
BDM signal samples are scaled to 10,000 events.
}
\label{fig:angular}
\end{figure}

Charged particles produced in the $\chi$-Ar interactions, as well as in the
propagation of neutral particles in liquid argon, are the observables in
LArTPCs~\cite{Abratenko:2019jqo,DUNETDRvol2}.
The four-momenta of the stable final-state particles, including protons,
electrons, photons, and charged pions,
are convolved with the detector resolution reported
in the DUNE Conceptual Design Report (CDR)~\cite{Acciarri:2015uup},
while the ones with kinetic energy below the detection threshold in DUNE CDR
and all the neutrons are excluded.
The distribution of $\cos\theta$,
where $\theta$ is the angle between the Sun's direction and the total momentum
of the final-state stable particles obtained from the procedure described above,
is shown in Fig.~\ref{fig:angular}.
In spite of the smearing from nuclear effects and detector resolution,
Fig.~\ref{fig:angular} quantitatively demonstrates that
the angular correlation in the BDM signal events persists, and can be exploited
to distinguish from the uniformly distributed background.
Selection criteria on $\cos\theta$ are optimized to different signal samples,
and efficiency ({\argonEfficiency}) and expected number of background events
({\expectedBackground}) in each
selection
are used to obtain the sensitivity of the BDM search.
Details of this analysis are described in the {\supplMat}.

\fakesubsection{Discussion}
\label{sec:discussion}
%
\begin{figure*}
\centering
\includegraphics[width=0.32\textwidth]{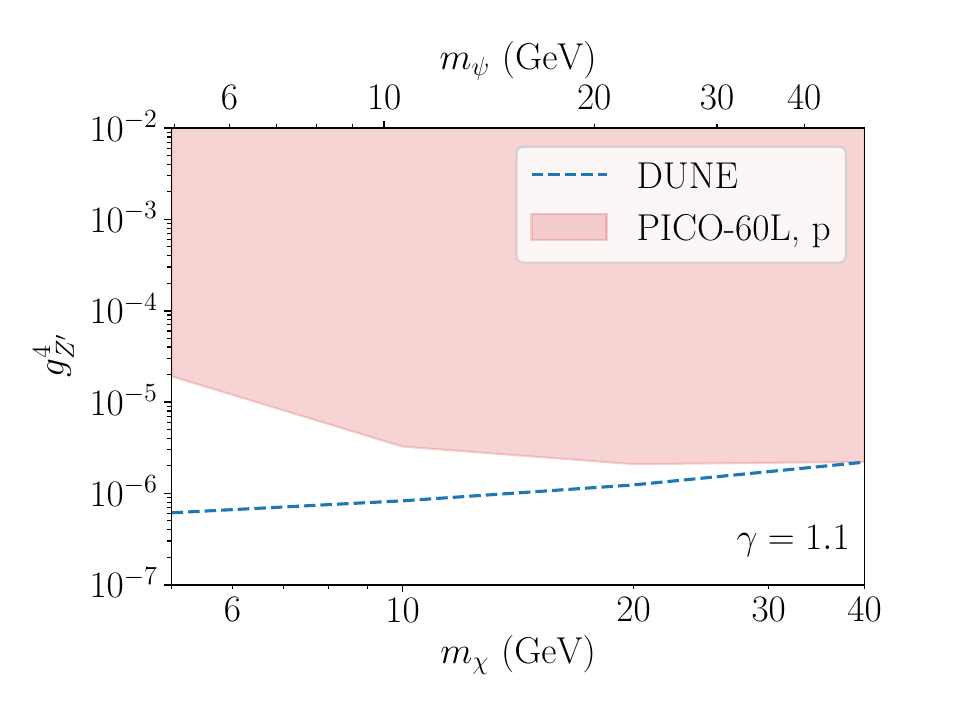}
\includegraphics[width=0.32\textwidth]{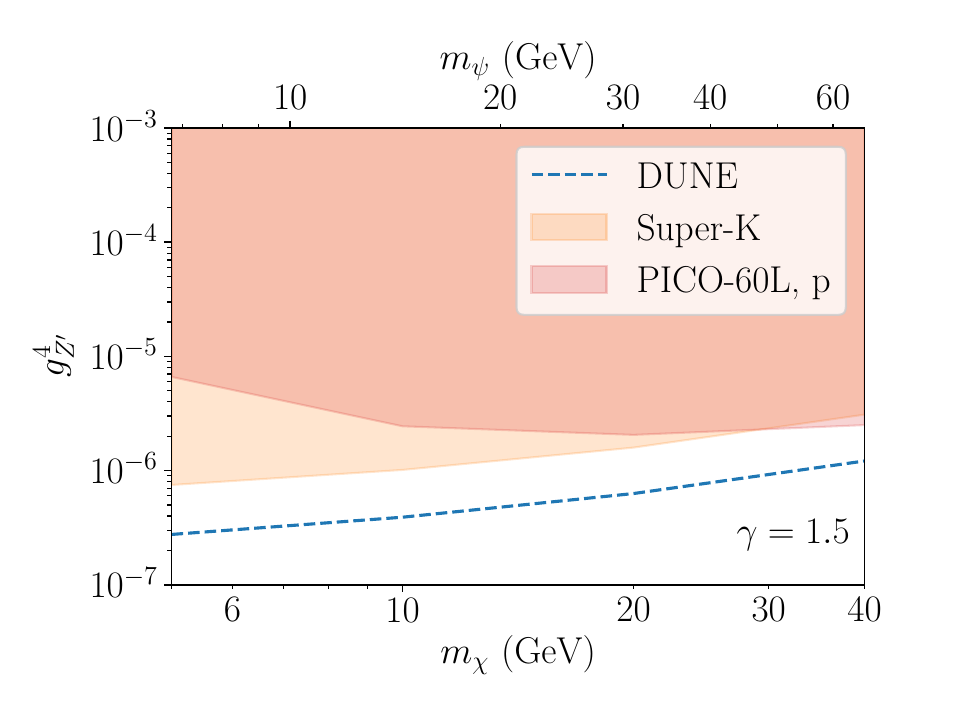}
\includegraphics[width=0.32\textwidth]{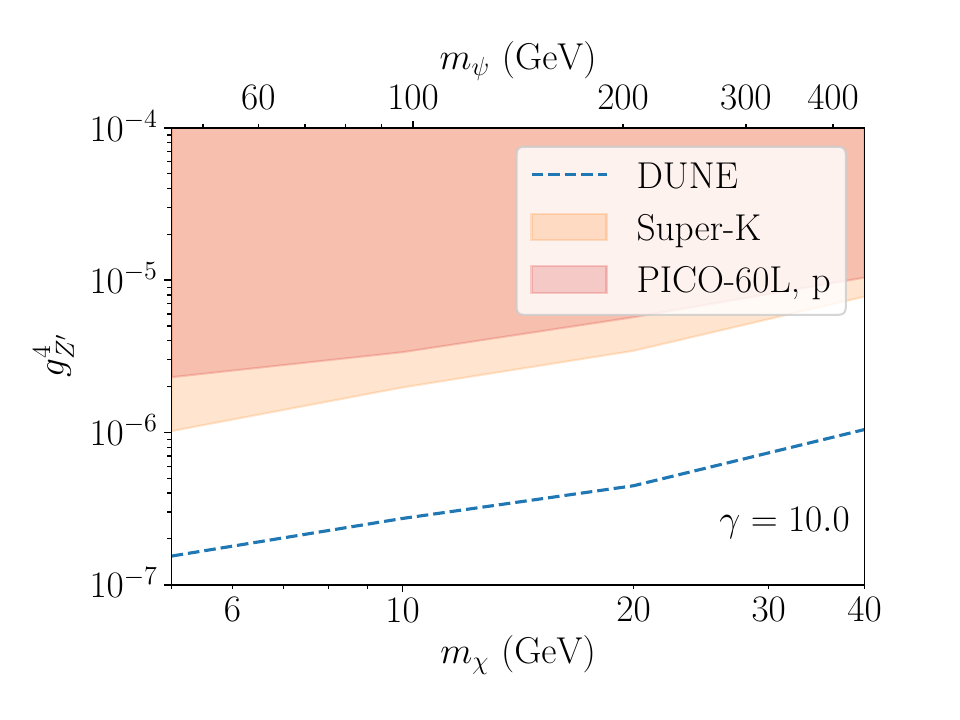}
\caption{Expected two standard deviation sensitivity with 10~years of livetime of a LArTPC with
fiducial mass of 40~kilotons, labeled as ``DUNE,'' 
for three
different benchmark DM boosts. The expected reach is compared to current
constraints from BDM ($\chi$) sensitivity with Super-Kamiokande and
spin-dependent direct detection searches for non-relativistic DM ($\psi$) at
PICO-60L~\cite{Amole:2017dex}. The leading constraints from direct detection of neutron
scattering by PandaX~\cite{Xia:2018qgs} are subdominant for the models we
consider.
\label{fig:significance}}
\end{figure*}
%
The projected sensitivity for 10 years of livetime with a DUNE-like detector with
40~kilotons of LAr is shown in Fig.~\ref{fig:significance}, in which the gauge coupling
constant in the benchmark model, $g_{Z^\prime}^4$, is excluded at two standard
deviations over the range of parameters considered.
Under the assumption that the $\chi$ relic abundance is negligible and
undetectable by direct detection experiments,
we compare the sensitivity of this benchmark model with the current constraint
at Super-Kamiokande by reinterpreting their atmospheric neutrino
measurement~\cite{Fechner:2009aa}, as detailed in the {\supplMat},
and spin-dependent direct detection searches for $\psi$~\cite{Amole:2017dex}.
Since Super-Kamiokande does not have sensitivity to BDM at $\gamma = 1.1$ due to
its high threshold for protons, it is absent in the first panel of
Fig.~\ref{fig:significance}.
It is worth noting that in a supporting study we also find that the fermionic BDM shows kinematic characteristics similar
to the scalar BDM,
and similar sensitivity can be achieved, while its parameter space is more
constrained by direct detection experiments~\cite{DUNETDRvol2}.

%
%

We demonstrate that underground, massive LArTPC detectors can have unique, complementary
capability of searching for BDM, taking into consideration the realistic nuclear
effects, detector resolution, and background for the first time.
Notably, we show that the sensitivity of this search technique is not compromised by the nuclear effects.
The method and the dedicated GENIE package we develop for this study offer the means to characterize
the BDM interactions in hadronic channels event by event,
and is straightforward to adapt to different models of nuclear effects
and atmospheric neutrino fluxes.
Our simulations offer the most accurate description of the signal and backgrounds for hadronic interactions to date.


BDM models are well-motivated and are gaining attention.
The framework of BDM~\cite{Fornal:2020npv,DelleRose:2020pbh,Ko:2020gdg,Alhazmi:2020fju,Chigusa:2020bgq}
has been explored for interpreting
the new observation from the Xenon1T experiment~\cite{Aprile:2020tmw}.
%
In that same context, this analysis pipeline will help narrow down the parameter space possible if a complementary observation occurs in LArTPC detectors;
it provides kinematic information to distinguish the signal across different classes
of BDM models, cross match with potential leptonic interactions also detectable with DUNE, and from neutrino detection of models of $\psi\overline{\psi}\to\nu\overline{\nu}$.
Additional studies that combine potential DUNE results with those from
direct detection or other experiments can help to further determine the properties of a BDM model.


We presented the first dedicated study of BDM search via hadronic interactions in underground,
massive LArTPCs, paving the avenue for future, sophisticated analyses.
The work can also be extended to investigate the requirements of detector specifications and reconstruction
criteria for BDM and similar astroparticle searches.



\begin{acknowledgments}
\fakesubsection{Acknowldgments}
We thank Costas Andreopoulos, Robert Hatcher, and Marco Roda for support related
to \GENIE.
We are grateful to Jonathan Asaadi, Mark Convery and Hirohisa Tanaka,
for all the discussion about the features of different detectors,
and Aaron Higuera, for the conversation regarding the rate of atmospheric neutrinos.
We also thank Jesse Thaler and Kaustubh Agashe for discussion.
JB is supported in part by PITT PACC. YC is supported in part by the US Department of
Energy under award number DE-SC0008541.
MG, GP, DS, and YTT are supported by the U.S. Department of Energy under
Contract No. DE-AC02-76SF00515.
LN is supported by the DOE under Award Number DESC0011632, and the Sherman Fairchild fellowship.
YZ is supported by U.S. Department of Energy under Award Number DESC0009959.
\end{acknowledgments}

\bibliography{bdm,DUNE}
\clearpage
\onecolumngrid
\setcounter{secnumdepth}{3}
\begin{center}
\large{\textbf{\title{\LetterTitle}}}\\  
\medskip
\supplMat
\end{center}
{Joshua~Berger, Yanou~Cui, Mathew~Graham, Lina~Necib,
Gianluca~Petrillo, Dane~Stocks, Yun-Tse~Tsai, and Yue~Zhao}
\vspace{0.5cm}\\
In this Supplementary Material we present certain details of our analysis that may be of interest to the reader, but are not essential to understanding our work. In Sec.~\ref{sup:theory}, we describe the physics entering the \GENIE BDM event generation module used in our analysis.
Sec.~\ref{sup:nuclear} discusses the effects of nuclear interactions on the observable particles.
We describe the background event generation procedures specific to our analysis in Sec.~\ref{sup:background}. Our detector simulation procedure is outlined in Sec.~\ref{sup:detsim}. Finally, in Sec.~\ref{sup:analysis} we describe our analysis strategy in detail.
\section{Boosted dark matter event generation (particle level)}
\label{sup:theory}

The liquid argon target is comprised, fundamentally, of electrons, quarks, and
gluons. Existing studies on BDM scattering so far have focused on BDM-electron
or BDM-nucleon (or BDM-H) scattering. A detailed study on BDM-nucleus scattering
is lacking and involves complex nuclear effects. In this section, we translate
quark-level interactions into cross-sections and event generation for
interactions of BDM with the nuclei in a target. These interactions are
implemented in the \GENIE Monte Carlo event generator and are now a
part of \GENIE version 3. The full details of these interactions and the
software package are presented in Ref.~\cite{Berger:2018}.

There are several regimes for this interaction depending on the kinematically
allowed momentum transfers for the interaction. As is standard with
neutrino-nucleus scattering, we parameterize the momentum transfer with
$Q^2=-q^2=-(k^\prime - k)^2$, where $k$ and $k^\prime$ are the four-momenta of
the incoming and outgoing BDM particle respectively. For an elastic scattering
on a nucleon at rest, one can relate this $Q^2$ to the outgoing nucleon kinetic
energy by $Q^2=2M_N\cdot E_{k,N}$, where $M_N$ is the nucleon mass and
$E_{k,N}$ is the outgoing nucleon kinetic energy.

At low momentum transfers ($Q^2 \ll (100~{\rm MeV})^2$), only coherent scattering off the nucleus is possible.
Since isotopes of argon with an odd number of neutrons are very rare and we are
considering models where spin-dependent interactions dominate, this process is
highly suppressed in argon and we neglect it entirely.

For $(100~{\rm MeV})^2 \lesssim Q^2 \lesssim (1~{\rm GeV})^2$, the only significant process is dark matter elastic scattering off of nucleons $\chi + N \to \chi + N$, where $N$ refers to a nucleon, that is a proton $p$ or a neutron $n$.
We will refer to this process simply as \emph{elastic scattering}, as is conventional in studying neutrino scattering.
In this regime, nuclear effects such as Fermi motion and Pauli blocking are relevant.
Furthermore, at higher momentum transfers, the nucleon form factor becomes an important effect.
All of these effects are described in detail below, in Sec.~\ref{sec:elastic}.

For $(800~{\rm MeV})^2 \lesssim Q^2 \lesssim (1.8~{\rm GeV})^2$, inelastic scattering begins to become an important process.
At these threshold momentum transfers, inelastic scattering is dominated by
resonant production of excited baryons $N^*$ and
$\Delta$, $\chi + N \to \chi + N^*$ and $\chi + N \to \chi + \Delta$.
This process, called \emph{resonant scattering}, is rather complicated to
describe and suffers from large modeling uncertainties.
In the present analysis, we omit these processes, rendering the limit projections we derive somewhat more conservative.
Description and modeling of these interactions will be performed in future work.

For $Q^2 \gtrsim (2~{\rm GeV})^2$, \emph{deep inelastic scattering} off partons in the nucleons, $\chi + q \to \chi + q$, becomes an increasingly good description.

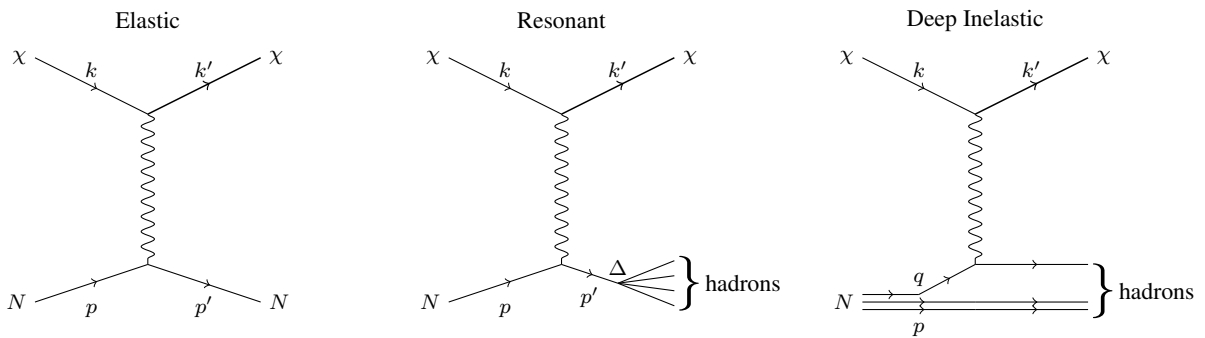
\begin{figure}[hbt]
\centering
\begin{tikzpicture}
\node at (0.5,0.5) {Elastic};
\draw[fermion] (-1,0) node[left] {\footnotesize $\chi$} -- node[above]
{\footnotesize $k$}
(0.5,-0.75);
\draw[fermion] (0.5,-2.75) -- node[below,text height=8pt] (pp)
{\footnotesize $p^\prime$} (2,-3.25)
node[right]
{\footnotesize $N$};
\draw[fermion] (-1,-3.25)  node[left] {\footnotesize $N$} --
node[below,baseline = (pp.base),text height=8pt] {\footnotesize $p$}
(0.5,-2.75);
\draw[fermion,line width=0.2mm] (0.5,-0.75) -- node[above]{\footnotesize
$k^\prime$} (2,0)
node[right]
{\footnotesize $\chi$};
\draw[gauge] (0.5,-0.75) -- (0.5,-2.75);

\node at (6,0.5) {Resonant};
\draw[fermion] (4.5,0) node[left] {\footnotesize $\chi$} -- node[above]
{\footnotesize $k$}
(6,-0.75);
\draw[fermion] (6,-2.75) -- node[below,text height=8pt]
{\footnotesize $p^\prime$} (6.75,-3)
node[above]
{\footnotesize $\Delta$};
\draw[fermion] (4.5,-3.25)  node[left] {\footnotesize $N$} --
node[below,baseline = (pp.base),text height=8pt] {\footnotesize $p$}
(6,-2.75);
\draw[fermion,line width=0.2mm] (6,-0.75) -- node[above]{\footnotesize
	$k^\prime$} (7.5,0)
node[right]
{\footnotesize $\chi$};
\draw(6.75,-3) -- (7.5,-2.7);
\draw(6.75,-3) -- (7.5,-2.9);
\draw(6.75,-3) -- (7.5,-3.1);
\draw(6.75,-3) -- (7.5,-3.3);
\node [anchor=west] at (7.4,-3) {\fontsize{22pt}{22pt}\selectfont\}};
\node[anchor=west] at (7.8,-3) {hadrons};
\draw[gauge] (6,-0.75) -- (6,-2.75);

\node at (11.5,0.5) {Deep Inelastic};
\draw[fermion] (10,0) node[left] {\footnotesize $\chi$} -- node[above]
{\footnotesize $k$}
(11.5,-0.75);
\draw[fermion] (10,-3.15) -- (10.75,-3.15) node[above] {\footnotesize $q$};
\draw[fermion] (10.75,-3.15) -- (11.5,-2.75);
\draw[fermion] (11.5,-2.75) -- (13,-2.75);
\draw[fermion] (10,-3.25)  node[left] {\footnotesize $N$} --
node[below,baseline = (pp.base),text height=8pt] {\footnotesize $p$}
(11.5,-3.25);
\draw[fermion] (11.5,-3.25) -- (13,-3.25);
\draw[fermion] (10,-3.35) -- (11.5,-3.35);
\draw[fermion] (11.5,-3.35) -- (13,-3.35);
\draw[fermion,line width=0.2mm] (11.5,-0.75) -- node[above]{\footnotesize
	$k^\prime$} (13,0)
node[right]
{\footnotesize $\chi$};
\node [anchor=west] at (12.9,-3.1) {\fontsize{22pt}{22pt}\selectfont\}};
\node[anchor=west] at (13.3,-3.1) {hadrons};
\draw[gauge] (11.5,-0.75) -- (11.5,-2.75);
\end{tikzpicture}

\caption{Diagrams illustrating each of the three processes that contribute to
DM scattering in argon.\label{fig:processes}}
\end{figure}
Diagrams illustrating a typical interaction for each of these processes are shown in Fig.~\ref{fig:processes}.
We now provide a detailed description of the elastic and deep inelastic scattering cross-sections and other relevant physics for each process.

Event generation for scattering in \GENIE proceeds via a series of modules that implement the relevant nuclear and particle physics.  Most of these implement nuclear physics effects, such as Fermi motion, Pauli Blocking and final state nuclear interactions.  These remain unchanged from their neutrino scattering implementation for BDM scattering.  We therefore focus below on the determination of the differential scattering cross-section as well as the BDM kinematics, which are the points at which BDM scattering differs from neutrino scattering.  We work here exclusively in the nucleon rest frame, which is not the same as the lab frame because of nucleon Fermi motion, but is reached by a trivial boost of the BDM-nucleon system.

\subsection{Elastic Scattering}
\label{sec:elastic}

The differential cross-section for elastic neutrino scattering in \GENIE
follows the calculation of Ahrens et.~al.~\cite{Ahrens:1986xe}, though the
formalism has been developed elsewhere in the literature and is standard.
As discussed in
the \emph{Letter}, we focus on the case where BDM interacts via a spin 1 boson
that has axial couplings to the quarks.  The amplitude for elastic scattering
then depends on hadronic matrix elements of the form:
\begin{equation}
\langle N | \overline{q_f} \gamma^\mu \gamma^5 q_f |N \rangle = \overline{u}
\left[F_A(Q^2) \gamma^\mu \gamma^5 + F_P(Q^2) \gamma^5 \frac{q^\mu}{m_N}\right]
u,
\end{equation}
where $F_A$ and $F_P$ are the axial and pseudo-scalar form factors
respectively, $q = k - k^\prime$ is the momentum transfer four-vector, and
$m_N$ is the nucleon mass.  For scalar BDM, it is straightforward to show that
the term
involving $F_P$ vanishes in the amplitude.  The differential cross-section in
the single kinematic variable $Q^2$
can be written as
\begin{equation}
\frac{d\sigma}{dQ^2} = \sigma_0 \left[A \pm B \frac{s - u}{m_N^2} + C
\left(\frac{s-u}{m_N^2}\right)^2\right],
\end{equation}
following the construction of Ref.~\cite{Ahrens:1986xe}.  The parameters are given by
\begin{eqnarray}
A & = & -Q_\chi^{2} \, \tau \, (\tau + \delta) \, (1 + \tau) \, |F_A|^2
\nonumber \\
B & = & 0 \nonumber \\
C & = & Q_\chi^{2} \, |F_A|^2.
\end{eqnarray}
with
\begin{equation}
\sigma_0 = \frac{g_{Z^\prime}^4 \, m_N^2}{4 \, \pi \, (E_\chi^2 - m_\chi^2) \,
(Q^2 + m_{Z^\prime}^2)^2},\qquad \tau = \frac{Q^2}{4\, m_N^2}, \qquad \delta
= \frac{m_\chi^2}{m_N^2}, \qquad
\frac{s - u}{m_N^2} = \frac{E_\chi}{m_N} - \tau.
\end{equation}
Here, $Q_\chi$ is the $Z^\prime$ charge of the scalar BDM and $s$, $u$ are
Mandelstam variables, and $E_\chi$ is the energy of the incident BDM.

The form factor $F_A$ is assumed to have a dipole form,
\begin{equation}
F_A(Q^2) \propto \frac{1}{(1 + Q^2 / M_A^2)^2},
\end{equation}
where $M_A$ is a parameter that needs to be fit to data.
The default value for this parameter in \GENIE, which we keep, is
$0.99~{\rm GeV}$.  The normalization of this form factor is given, in general,
by a combination of the spin form factors of the nucleon.
Assuming isospin symmetry, the form factors
for the proton and neutron are
\begin{equation}
F_A^p(0) = Q_u \Delta u + Q_d \Delta d + Q_s \Delta s,\qquad F_A^n(0) = Q_u
\Delta d + Q_d \Delta u + Q_s \Delta s,
\end{equation}
with the quark axial charges $Q_{f}$.
The spin form factors need to be either extracted from data or calculated on the lattice.  We take them to be~\cite{Alexandrou:2017hac}
\begin{equation}
\Delta u = 0.84,\qquad \Delta d = -0.43,\qquad \Delta s = -0.09.
\end{equation}

Note that the range of momentum transfers is given by
\begin{equation}
0 < Q^2 < 4 \frac{m_N^2 (E_\chi^2 -\BDMmass^2)}{\BDMmass^2 + 2 E_\chi m_N +
m_N^2}.
\end{equation}

\subsection{Deep Inelastic Scattering}
\label{sec:deep-inelastic}

The phase space for deep inelastic scattering (DIS) is described by two, rather
than one, variables, in addition to the complex hadronic phase space determined
by the hadronization procedure.
One intuitive way of breaking down the phase space here
is in terms of the momentum transfer $Q^2$ and the total invariant mass of the
final state hadronic system $W$.

While the variables $Q^2$ and $W$ are physically intuitive, it is simpler to
describe the cross-section in terms of variables $x$ and $y$, where $x$ is the
usual Bjorken variable
and $y$ is the fractional energy loss of the
incoming DM particle,
\begin{equation}
y = 1 - \frac{E_\chi^\prime}{E_\chi},
\end{equation}
where $E_\chi$ and $E_\chi^\prime$ are the energy of incoming and outgoing DM,
respectively.
These variables can be written in Lorentz invariant form, related to $Q^2$ and $W^2$ as
\begin{equation}
x = \frac{Q^2}{Q^2 + W^2 - m_N^2}\, \qquad y = \frac{Q^2 + W^2 - m_N^2}{2 \, E_\chi \, m_N}.
\end{equation}
Note that these variables range in a subset of $0 < x,y < 1$ that can be solved for numerically.

To proceed and calculate the cross-section, we follow closely the notation of Ref.~\cite{Paschos:2001np}.
We define the hadronic tensor as the initial spin averaged, final state summed squared hadronic matrix element at fixed $Q^2$ and $W^2$, summed and integrated overall all possible final states.  By Lorentz invariance, the hadronic tensor has the form
\begin{equation}
W^{\mu\nu} = - g^{\mu\nu} \, F_1(x, Q^2) + \frac{p^\mu \, p^\nu}{p \cdot q} \,
F_2(x, Q^2) - i \epsilon^{\mu\nu\rho\sigma} \frac{p_{\rho} q_\sigma}{2 \, p
\cdot q} \, F_3(x, Q^2) + \frac{q^\mu \, q^\nu}{p \cdot q} F_4(x, Q^2) +
\frac{p^\mu \, q^\nu + q^\mu \, p^\nu}{2 \, p \cdot q} F_5(x, Q^2),
\end{equation}
where $p$ is the four-momentum of the initial nucleon. The $F_i$ are structure functions that are related to the quark PDFs
below.
For scalar DM scattering, we find
\begin{equation}
\frac{d\sigma}{dx \, dy} = \frac{g_{Z^\prime}^4 \, m_N \, E_\chi^3}{32 \, \pi
\, (E_\chi^2 - \BDMmass^2)} \, \left[-4 \, Q_\chi^{2} \, y \, \left(x \, y + 2
\frac{\BDMmass^2}{m_N \, E_\chi}\right) F_1 + 2 \, Q_\chi^{2} \, (y - 2)^2 \,
F_2\right].
\end{equation}
The structure functions $F_i$ here are given in terms of the quark PDFs by the following relations by
\begin{eqnarray}
F_2 & = & 4 \, x \, \sum_f (Q_f^{2}) \, [f_f(x, Q^2) + f_{\bar{f}}(x , Q^2)]
\nonumber \\
F_3 & = & 0,
\end{eqnarray}
where $f_f$ are the parton distribution functions for quark flavor $f$, combined with the Callan-Gross relation
\begin{equation}
2 \, x \, F_1 = F_2,
\end{equation}
and the Albright-Jarlskog relations
\begin{equation}
F_4 = 0, \qquad x\, F_5 = F_2.
\end{equation}
The parton distributions used in \GENIE are a patched version of the GRV98lo PDFs~\cite{Gluck:1998xa}.

Once the $x$ and $y$ of a DIS event are selected, the hadronic final state phase space is then populated using one of two hadronization models.  At low energies, an empirical Koba-Nielson-Olesen (KNO) model is used in the neutrino.  Absent empirical observation of BDM, we must make an assumption of the empirical behavior to implement this model for DM.  We assume that BDM scattering behaves like neutrino scattering in the KNO model.  At high energies, \PYTHIA is used to hadronize the final state hadronic system.  This procedure remains unchanged for BDM scattering.

For details of DIS interactions for fermionic BDM, see \cite{Berger:2018}.

\subsection{ Resonant scattering}

Resonant scattering via excited baryon states is implemented for neutrino scattering in \GENIE, but the implementation of their models for BDM is challenging to validate.  This process is not studied in the present analysis, though it can only increase the sensitivity to BDM.

\section{Impact from nuclear effects}
\label{sup:nuclear}

In the recent years,
interest in the interactions of hadrons produced within the nucleus on their
way out of the nuclear remnant
(``final state interactions'') has surged within the community
owing to their significant impact on precision measurements of neutrino oscillations
and search for nucleon decays,
especially with detectors based on large nuclei like oxygen and argon.

In this analysis, the default $hA$ final
state interaction (FSI) model in \GENIE~
\cite{Merenyi:1992gf,Ransome:2005vb,Andreopoulos:2015wxa} is used to
model the nuclear FSI, but it is
straightforward to switch to different models in the future data analysis with
the tool we developed for this study. This
model uses empirically determined total cross
sections for various processes that
hadrons propagating through the nuclear remnant can undergo, such as pion
absorption, elastic and inelastic scattering, and charge exchange. The cross
sections are extrapolated to high energies where data is unavailable. The focus
of this model was on iron for the MINOS experiment. Alternate models currently
include the $hN$ model, which implements a more complex intranuclear cascade
designed for situations with multiple scattering. On the other hand, it
currently does not include important medium corrections~\cite{Dytman:2015taa}.
More recent iterations of both models have been developed as well.

In addition to FSI, there is some modification of the kinematics due to nucleon
motion within the nucleus and Pauli blocking. \GENIE
models these phenomena with a Fermi gas.

\begin{figure}[!htb]
\centering
\includegraphics[width=0.5\textwidth]{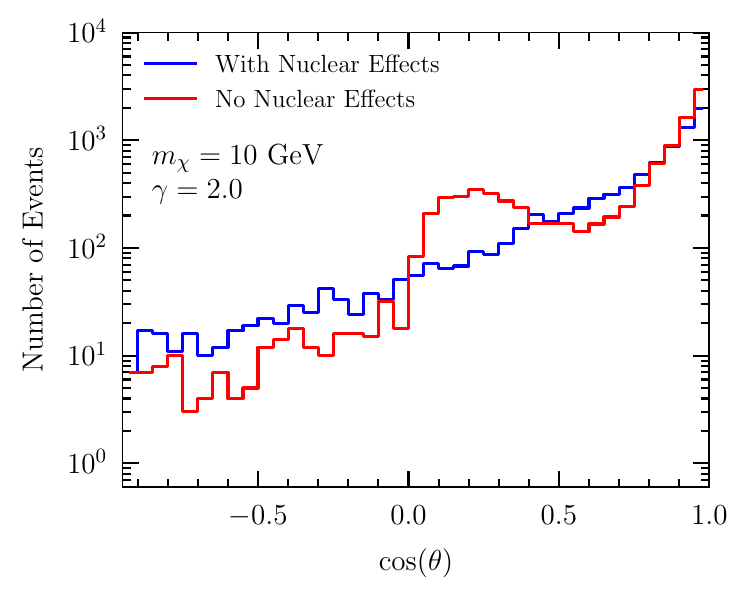}
\caption{The impact of the nuclear effects on the distribution of $\cos\theta$,
the angle between the total momentum of the final state visible particles
(excluding neutrons and the outgoing BDM) and the incident BDM.}.
\label{fig:NuclearEffect}
\end{figure}

Figure~\ref{fig:NuclearEffect} illustrates
the impact of the nuclear effects on the distribution of $\cos\theta$,
where $\theta$ is the angle between the total momentum of the final state visible
particles (i.e. excluding neutrons and the outgoing BDM) and the incident BDM.
The kinematic feature at $\cos\theta \approx 0.25$, which
originates from the elastic component of the scattering,
gets smeared out dominantly by the effects
of nucleon Fermi motion, which misaligns the detector and nucleon rest frames.

\section{Background}
\label{sup:background}

As stated in the \emph{Letter}, we consider the neutral current interactions of
neutrinos produced in the atmosphere as the main background.
The absolute rate of interactions from atmospheric neutrinos is calculated
integrating the neutrino cross sections with argon with the atmospheric neutrino fluxes
using Bartol model~\cite{PhysRevD.70.023006} at the MINOS location in Soudan,
as in~\cite{DUNETDRvol2} p.~6-193.
The simulation of the background processes is performed using the \LArSoft
toolkit~\cite{Snider:2017wjd} interfaced with \GENIE,
including generation of $\nu_{\mu}$, $\bar{\nu}_{\mu}$,
$\nu_{e}$ and $\bar{\nu}_{e}$ distributed according to the reference Bartol
flux.
Additional corrections to account for a larger energy spectrum
and for neutrino flavor oscillations are described below.

As with the signal, each generated neutrino interaction is assigned
a direction toward the Sun randomly extracted
from the unbiased distribution of the Sun position with respect to the
DUNE far detector. This direction is solely used to estimate the angle
$\cos \theta$ used as an observable in this study.

\subsection{Solar magnetic activity}

Solar magnetic activity affects cosmic ray deflection and as a consequence
the rate and spectrum of atmospheric neutrinos.
The activity oscillates between minimal and maximal with a period of about 11
years. Atmospheric neutrino fluxes include the effects of this activity,
and are provided separately in the minimum and maximum activity scenarios.
In our study we mix for each process two samples simulated with the two
scenarios. Since the period of this activity is close to the 10 years of
duration of data taking we consider in this study, we assume one full cycle
and therefore we mix the two samples with equal weight.

\subsection{Extension of background estimation to higher neutrino energy}

Our atmospheric neutrino background estimation
includes only neutrino energies between 100\,MeV and 10\,GeV,
being based on Bartol flux at Soudan~\cite{BartolFluxURL}.
With our choice of parameters, BDM interactions can cover a larger energy range
and we need to extend the background estimation to cover that range;
because the neutrino flux rapidly decreases with the neutrino energy $E_{\nu}$,
roughly as $E_{\nu}^{-2}$, we elect to extend the coverage only up to 100\,GeV.
To do so, we use the Honda flux (at Homestake),
which extends up to 10\,TeV,
scaling it so that the flux integrated in the energy range 1 to 10\,GeV
matches DUNE background estimation.
The approximation implied by this procedure is that the two atmospheric neutrino models,
Bartol and Honda, scale with energy in the same way.
We estimate this approximation to carry an error of about 20\%.
The choice not to employ Bartol fluxes for this extension is purely technical,
due to a temporary issue in the 
\LArSoft software.
Likewise, the choice to use samples with narrow energy ranges is technical.
Due to the steep decrease of flux with energy,
the generation of a single sample with large energy range uses computational
resources very inefficiently.

High energy neutrino events inherently present kinematics different from
lower energy ones. In our simple analysis events are selected according to a
single quantity: the angle ($\cos \theta$) between the direction of the
reconstructed particles and direction of the Sun. These two directions are
uncorrelated for the atmospheric neutrino background and $\cos \theta$ is
mostly independent from the neutrino energy. We confirm that the Honda
high energy neutrino NC interaction sample shows the same distribution in
$\cos\theta$ as the Bartol atmospheric neutrino sample that constitutes our
reference background (see Fig.~\ref{fig:angular})
and that both $\cos\theta$ distributions are consistently uniform.
Because of this, we simply retain the reference background sample in this analysis,
scaling its size up by a factor to account for the high energy contribution.
The small size of the resulting correction, 3.8\%,
suggests that interactions with even higher energy neutrinos above 100\,GeV
will contribute negligibly to this background.

\subsection{Tau neutrino background}

Under the assumption of being able to identify and discard background events
where charged-current interactions produce electrons or muons,
our background is constituted mainly of atmospheric neutrino interactions via neutral current.
An exception is a charged-current interaction where a $\tau$ lepton is produced
that decays into a neutrino and hadrons. This happens with a branching
fraction close to 60\%.
While neutrinos of $\tau$ flavor are rarely produced in the interaction of
cosmic rays with the atmosphere,
it is still possible for a muon neutrino to transform (``oscillate'') into $\nu_{\tau}$.
The probability of this transition for a muon neutrino of energy $E_{\nu}$ is described by the formula
\begin{equation}
P\left(\nu_{\mu} \rightarrow \nu_{\tau}\right) \approx \cos^{4} \theta_{13}\
\sin^{2} 2\theta_{23}\ \sin^{2} \frac{\Delta m^{2}_{31} L}{4 E_{\nu}},
\end{equation}
where the parameters $\theta_{13}$, $\theta_{23}$ and $\Delta m^{2}_{31}$ have
been measured \cite{Tanabashi:2018oca}.
An accurate computation of the rate is complicated by the dependency on $L$,
the distance from the point in the atmosphere where the neutrino is produced to the point in the detector where it interacts.
This distance can be as short as a few kilometers for neutrinos produced right above the detector,
to more than ten thousand kilometers for the ones produced at the opposite side of the Earth;
this is compared to the factor $\Delta m^{2}_{31}/ 4 \approx
3\,\textnormal{MeV}\!/\textnormal{km}$.
We simplify the problem by the very conservative approximation of oscillation probability
being maximal independently from $E_{\nu}$,
by setting the last of the three factors of the expression above to $1$,
yielding $ P\left(\nu_{\mu} \rightarrow \nu_{\tau}\right) \approx 95\%$,
with the understanding that this represents a significant overestimation of
this component of the background.

Oscillation does not have observable consequences on neutral-current interaction backgrounds.

We generate the tau neutrino sample using the same Honda flux as for muon
neutrinos.
To ensure that the size of the $\nu_{\tau}$ samples is consistent with the other background samples,
we impose the same rate of interaction via \emph{neutral} current for
$\nu_{\mu}$ and $\nu_{\tau}$ of the same energy, by properly scaling the tau
neutrino interaction rate.
The rate of interaction of $\nu_{\tau}$ via charged current,
the one relevant for this part of the background,
is scaled with the same factor,
but it remains much smaller than for $\nu_{\mu}$ at low energy,
being suppressed by the larger mass of $\tau$ lepton.
The charged-current $\nu_{\tau}$ interactions are mostly suppressed for
$E_{\nu} < 10\,{\rm GeV}$,
whereas the charged-current $\nu_{\mu}$ interactions with the same energy range,
$E_{\nu} < 10\,{\rm GeV}$,
constitutes 90\% of those in the full energy range we consider,
$100\,{\rm MeV} < E_{\nu} < 100\,{\rm GeV}$.
For this reason, the $\nu_{\tau}$ charged-current background is much smaller
than the atmospheric neutrino neutral-current background,
2.8\% in our estimation.
As for the contribution to the background from the high energy extension,
the reference background sample, i.e.\ the atmospheric neutrino sample based on Bartol flux
and with no oscillation, is still used for the analysis,
and the effect of oscillation into $\nu_{\tau}$ is included
as a 2.8\% correction factor on the total background rate.

\section{Detector simulation}
\label{sup:detsim}

For each event,
we use \GEANT~\cite{AGOSTINELLI2003250,1610988,ALLISON2016186} to simulate the propagation of the final-state SM
particles in liquid argon until any short-lived particles have decayed.
The four-momentum of the stable particles,
protons, neutrons, charged pions, muons, electrons, and photons, are
convolved based on the parameters characterizing the detector response.
The convolution of the four-momenta accounts for the energy
and angular resolution of the detector.
Only the particles with their convolved energy greater than the detector
threshold are taken into account in the subsequent steps of the analysis.
The baseline scenario deployed in this analysis consists of
the detector response and threshold reported in the DUNE Conceptual Design
Report (CDR)~\cite{Acciarri:2015uup}, as listed in Table~\ref{tab:det_resolution},
and, of no neutron detection.
The resulting sensitivity on BDM search is presented in the \emph{Letter}.

\begin{table}[!htb]
\centering
\begin{tabular}{c@{$\quad$}c@{$\quad$}l@{$\quad$}c}
\hline\hline
Particle type & Detection Threshold (KE) & Energy Resolution & Angular Resolution \\
\hline
$\mu^{\pm}$ & 30 MeV &  5\% & $1^{\circ}$ \\
$\pi^{\pm}$ & 100 MeV & 5\% & $1^{\circ}$ \\
e$^{\pm}$/$\gamma$ & 30 MeV & $2\%\oplus 15\%/\sqrt{E}$~[GeV] & $1^{\circ}$ \\
p & 50 MeV & $p<400~{\rm MeV/c}$: 10\%  & $5^{\circ}$ \\
  &        & $p>400~{\rm MeV/c}$: $5\%\oplus 30\%/\sqrt{E}$~[GeV] &  \\
n & 50 MeV & $40\%/\sqrt{E}$~[GeV] & $5^{\circ}$ \\
\hline\hline
\end{tabular}
\caption{Summary of the detection threshold in kinetic energy (KE) and
the detector response, including the energy and
angular resolution, for stable particles from DUNE CDR~\cite{Acciarri:2015uup}.}
\label{tab:det_resolution}
\end{table}

To evaluate the impact from the detector response and threshold, as well as
the capability of reconstructing neutrons,
an alternative set of energy resolution is deployed.
This set of tabulated energy resolution was obtained and studied by the authors
of Ref.~\cite{PhysRevD.99.036009}.
In addition, we study the cases where 90\% of neutrons can be detected and
reconstructed.
Further, we lower the detection threshold to 20~MeV in kinetic energy (KE)
for protons, neutrons, and to 30~MeV in KE for charged pions,
labeled as the ``optimistic'' scenario for detection thresholds.
All the scenarios are outlined in Table~\ref{tab:scenarios}.

\begin{table}
\centering
\begin{tabular}{c@{$\quad$}c@{$\quad$}c@{$\quad$}c@{$\quad$}c}
\hline\hline
Scenario & Energy Resolution & Angular Resolution & Neutron Efficiency & Detection Threshold \\
\hline
1 & DUNE CDR & DUNE CDR & 90\% & DUNE CDR \\
2 & DUNE CDR & DUNE CDR & 0\%  & DUNE CDR \\
3 & Ref.~\cite{PhysRevD.99.036009} & DUNE CDR & 90\% & DUNE CDR \\
4 & Ref.~\cite{PhysRevD.99.036009} & DUNE CDR &  0\% & DUNE CDR \\
5 & DUNE CDR & DUNE CDR & 90\% & Optimistic \\
6 & DUNE CDR & DUNE CDR &  0\% & Optimistic \\
\hline\hline
\end{tabular}
\caption{Summary of the different scenarios on the detector response,
threshold, and neutron reconstruction efficiency studied in this analysis.
The final results reported in the \emph{Letter} are based on the first (baseline)
scenario, which incorporates the detector response reported in DUNE
CDR~\cite{Acciarri:2015uup} and is summarized in Table~\ref{tab:det_resolution}.
}
\label{tab:scenarios}
\end{table}

We obtain similar sensitivity on BDM search from all the scenarios being tested.
This is owing to the fact that we deploy a simple analysis approach,
as depicted in Section~\ref{sup:analysis},
and do not utilize plenty of information to which better energy resolution,
lower detection threshold, or capability of neutron reconstruction is
relevant.

\section{Analysis}
\label{sup:analysis}

The BDM signal events are expected to have final-state particles roughly aligned
with the incoming BDM particle, which we take to be coming from the Sun.
We use this feature to select events with enhanced the signal-to-background
ratio.

\subsection{Baseline Analysis}
\label{sup:baseline_analysis}

We develop selection criteria based on $\theta$, which,
as defined in the \emph{Letter},
is the angle between the total momentum of the final-state stable SM particles
and the incident BDM (aligned with the Sun).
The detector response and thresholds are taken into account,
as described in Sec.~\ref{sup:detsim}.
The single variate selection
is optimized to the minimal signal strength, $s^{\prime}$,
for which we could obtain a sensitivity to BDM signal at 5 standard deviations,
\begin{equation}
\frac{\argonEfficiency s^{\prime}}{\sqrt{\argonEfficiency s^{\prime} + \expectedBackground}} = 5.
\label{eq:optimization}
\end{equation}
The factor \argonEfficiency{} represents effectively the product of the acceptance and efficiency
of the signal selection,
while the expected number of selected BDM events,
{\expectedBDMsignal}, can be written as
$\expectedBDMsignal = \argonEfficiency s^{\prime}$.
We evaluate \argonEfficiency{} and the number of the
background events \expectedBackground{} respectively
from the BDM and atmospheric neutrino MC samples, and
the selection criterion on $\cos\theta$ is individually optimized to each
benchmark BDM signal sample, as tabulated in Table~\ref{tab:eff-tab}.
Note that Eq.~\ref{eq:optimization} is used for optimizing the selection
criteria, but not for extracting the sensitivity to the BDM signal.

\begin{table*}[!t]
\centering
\begin{tabular}{c@{$\quad$}c@{$\quad$}c@{$\quad$}c@{$\quad$}c@{$\quad$}c@{$\quad$}c}
\hline\hline
\BDMmass (GeV) & $\gamma$ & \argonEfficiency & $\referenceXsecBDMAr / g_{Z^\prime}^4~({\rm cm}^{2})$  & $\referenceBDMflux/g_{Z^\prime}^4~({\rm cm}^{-2} {\rm s}^{-1})$ & $\expectedBDMsignal/g_{Z^\prime}^8$ & \expectedBackground \\
\hline
5  & 1.1  & $0.4917$ & $9.057 \times 10^{-30} $ & $634.1$ & $5.32 \times 10^{14}$ & $10006\pm113$ \\
10 & 1.1  & $0.4788$ & $1.063 \times 10^{-29} $ & $303.6$ & $2.91 \times 10^{14}$ & $10006\pm122$ \\
20 & 1.1  & $0.4973$ & $1.220 \times 10^{-29} $ & $117.7$ & $1.34 \times 10^{14}$ & $10634\pm126$ \\
40 & 1.1  & $0.5027$ & $1.278 \times 10^{-29} $ & $36.38$ & $4.40 \times 10^{13}$ & $11300\pm130$ \\ 
5  & 1.5 & $0.6532$ & $4.978 \times 10^{-29} $ & $468.3$ & $2.87 \times 10^{15}$ & $11894\pm133$ \\
10 & 1.5 & $0.6660$ & $5.609 \times 10^{-29} $ & $203.4$ & $1.43 \times 10^{15}$ & $11894\pm133$ \\
20 & 1.5 & $0.6752$ & $5.965 \times 10^{-29} $ & $72.48$ & $5.50 \times 10^{14}$ & $11894\pm133$ \\
40 & 1.5 & $0.6694$ & $6.152 \times 10^{-29} $ & $19.10$ & $1.482 \times 10^{14}$ & $11894\pm133$ \\
5  & 10   & $0.7635$ & $1.270 \times 10^{-27} $ & $28.12$ & $5.13 \times 10^{15}$ & $ 3723\pm 74$ \\
10 & 10   & $0.7673$ & $1.377 \times 10^{-27} $ & $7.521$ & $1.50 \times 10^{15}$ & $ 3075\pm 68$ \\
20 & 10   & $0.8366$ & $1.437 \times 10^{-27} $ & $2.455$ & $5.56 \times 10^{14}$ & $ 3075\pm 68$ \\
40 & 10   & $0.8512$ & $1.470 \times 10^{-27} $ & $0.431$ & $1.02 \times 10^{14}$ & $ 3075\pm 68$ \\
\hline\hline
\end{tabular}
\caption{
  Efficiency to detect DM recoils in a LArTPC (\argonEfficiency),
  the BDM-Ar cross-section (\referenceXsecBDMAr),
  the expected solar-produced BDM flux (\referenceBDMflux),
  and expected number of signal (\expectedBDMsignal) and background events (\expectedBackground) after cuts
  for our benchmark models (mass \BDMmass{} and boost $\gamma$)
  assuming an exposure of 40~\kt and 10~years.
  \label{tab:eff-tab}
  }
\end{table*}

\subsection{Alternative Selection Oriented to Moderate Boost Signals}
\label{sup:alternative_analysis}

Owing to the kinematics of the elastic scattering,
the $\cos\theta$ distributions in the moderate boost signal samples
(e.g.~$\gamma = 1.1$)
are more widely spread, and, as a consequence,
the single variate selection based on $\cos\theta$ is less efficient,
resulting in a smaller
{\argonEfficiency} and a greater {\expectedBackground} in Table~\ref{tab:eff-tab}.
To improve the signal-to-background ratio in these cases,
the kinematic correlation between $\cos\theta$ and $P$ is studied,
where $P$ denotes the value of the total three-momentum of the final-state
SM particles. In the limit that the nucleon is at rest when it is struck by
the incident BDM, these variables are perfectly correlated for a given model
at fixed invariant mass for the final-state hadronic system. This correlation
does not hold for the background and should allow for further separation of
signal and background.
A few cut-based analyses using the two variables, $\cos\theta$ and $P$,
are explored.
With the simple statistic estimate used in this analysis,
the sensitivity is comparable to the baseline analysis;
however,
we expect more sophisticated analyses, with better understanding on
nuclear effects and detector response,
to significantly improve the sensitivity of the BDM search.

\subsection{Impacts from detector response and threshold}
\label{sup:impact_detsim}

We study the impacts from different scenarios of detector response and
threshold by performing the baseline analysis with the convolved four-momentum
from all the scenarios listed in Table~\ref{tab:scenarios}.
In addition, we compare the results combining different detector response
and analysis strategies (baseline analysis versus alternative analysis).
Similar to the conclusion obtained from Sec.~\ref{sup:alternative_analysis},
to significantly improve the sensitivity of BDM search
requires better understanding on the BDM signal and atmospheric neutrino
background, including nuclear effects and the flux of atmospheric neutrinos,
as well as more sophisticated analyses.

\subsection{Statistical Method}
\label{sup:statistical_method}

We obtain the projected sensitivity for 10 years of livetime with a DUNE-like detector with
40~\kt of LAr.
Since we expect a large number of signal events
for the parameter space at the boundary of the discovery reach,
the expected significance is evaluated with a large statistics estimate~\cite{Tanabashi:2018oca},
\begin{equation}
  Z \approx \sqrt{2 \left[(\expectedBDMsignal+\expectedBackground) \log\left(1 + \frac{\expectedBDMsignal}{\expectedBackground}\right) - \expectedBDMsignal\right]},
\end{equation}
where {\expectedBDMsignal} and {\expectedBackground} are the numbers of
expected signal and background events respectively.  We find that
our LArTPC reference detector is
sensitive to $g_{Z^\prime}^4 = (1.54-22.0) \times 10^{-7}$ at two standard
deviations over the range of parameters considered, as shown
in Fig.~\ref{fig:significance}.

\subsection{Super-Kamiokande Data Comparison}
\label{sup:superK}

A reinterpretation of the NC elastic $\nu+p\to\nu+p$ measurement from the
atmospheric neutrino events collected in Super-Kamiokande is performed for
comparison with our LArTPC analysis. The BDM events scattered on hydrogen and
oxygen atoms in  Super-Kamiokande are simulated by the same BDM module in
\GENIE, and, accounting
for the \Cherenkov threshold and the \Cherenkov cone selection and efficiency
in \cite{Fechner:2009aa}, the events containing a single proton
with momentum between 1.07 and 2.62~GeV are selected and scaled.
The sensitivity of the BDM signals in Super-Kamiokande is thereby evaluated
based on the simulated BDM events and the atmospheric neutrino data
in Tables I and II in \cite{Fechner:2009aa}.


\end{document}